\documentclass[aps,prb,twocolumn,groupedaddress]{revtex4}

\usepackage{graphicx}
\usepackage{dcolumn}
\usepackage{bm}
\usepackage{subfigure}

\begin{document}


\title{FINITE-SIZE SCALING OF THE DOMAIN WALL ENTROPY FOR THE 2D
$\pm J$ ISING SPIN GLASS}


\author{Ronald Fisch}
\email[]{ron@princeton.edu}
\affiliation{382 Willowbrook Dr.\\
North Brunswick, NJ 08902}


\date{\today}

\begin{abstract}
The statistics of domain walls for ground states of the 2D Ising
spin glass with +1 and -1 bonds are studied for $L \times L$
square lattices with $L \le 20$, and $x$ = 0.25 and 0.5, where $x$
is the fraction of negative bonds, using periodic and/or
antiperiodic boundary conditions. Under these conditions, almost
all domain walls have an energy $E_{dw}$ equal to 0 or 4.  The
probability distribution of the entropy, $S_{dw}$, is found to
depend strongly on $E_{dw}$.  The results for $S_{dw}$ when
$E_{dw} = 4$ agree with the prediction of the droplet model. Our
results for $S_{dw}$ when $E_{dw} = 0$ agree with those of Saul
and Kardar. In addition, we find that the distributions do not
appear to be Gaussian in that case. The special role of $E_{dw} =
0$ domain walls is discussed, and the discrepancy between the
prediction of Amoruso, Hartmann, Hastings and Moore and the result
of Saul and Kardar is explained.

\end{abstract}


\maketitle

\section{INTRODUCTION}

There continues to be a controversy about the nature of the Ising
spin glass.  The Sherrington-Kirkpatrick model,\cite{SK75} with
its infinite-range interactions between the spins, is described by
the Parisi replica-symmetry breaking mean-field
theory.\cite{Par80,Par83} To understand models with short-range
interactions on finite-dimensional lattices, however, it is
necessary to include the effects of interfaces, which do not exist
in a well-defined way in an infinite-range model.  The droplet
model of Fisher and Huse,\cite{FH86,HF86,FH88} which starts from
the domain-wall renormalization group ideas of
McMillan\cite{McM84a,McM84b,McM85} and Bray and Moore,\cite{BM85,
BM86} and studies the properties of interfaces, provides a very
different viewpoint on the spin-glass phase.  More recently,
numerical results\cite{KM00,PY00,HG02} indicate that the actual
situation for three-dimensional (3D) models combines elements of
both pictures in a nontrivial way.

In two dimensions (2D), the spin-glass phase is not stable at
finite temperature.  Because of this, it is necessary to treat
cases with continuous distributions of energies (CDE) and cases
with quantized distributions of energies (QDE)
separately.\cite{BM86,AMMP03}

In three or more space dimensions, where a spin-glass phase is
believed to occur at finite temperature, the general framework of
thermodynamics requires that the CDE and the QDE should be treated
on the same footing. The way this comes about is that in these
cases the typical domain wall energy increases as a positive power
of the size of the lattice. Thus the quantization energy becomes a
negligible fraction of the domain wall energy for large lattices.
All bond distributions behave in a qualitatively similar way,
except that the QDE have finite ground state
entropies.\cite{BM86,FH88}

Amoruso, Hartmann, Hastings and Moore\cite{AHHM06} have recently
proposed that in 2D there is a relation
\begin{equation}
  d_f = 1 + {3 \over {4 ( 3 + \theta_E )}}   \, ,
\end{equation}
where $d_f$ is the fractal dimension of domain walls, and
$\theta_E$ is the exponent which characterizes the scaling of the
domain wall energy with size.  For continuous energy distributions
the existing numerical estimates of $d_f$ and $\theta_E$ satisfy
Eqn. (1).

It is known from the droplet theory,\cite{FH88} that for the QDE,
which have a positive entropy at zero temperature,
\begin{equation}
  d_f = 2 \theta_S   \, .
\end{equation}
$\theta_S$ is the exponent for the scaling of domain wall entropy
with size.  Thus, for the QDE, Eqn. (1) provides a relation
between the scaling of the energy and the entropy of domain walls.
It is not known how to calculate $d_f$ directly for the QDE case,
so we need to use Eqn. (2) to verify Eqn. (1) in that case.

For the QDE, it is known that $\theta_E = 0$.\cite{AMMP03,HY01}
Then using Eqn. (1) gives $d_f = 5/4$, or using Eqn. (2),
$\theta_S = 5/8$. The calculation of $\theta_S$ by Saul and
Kardar,\cite{SK93,SK94} found $\theta_S = 0.49 \pm 0.02$.  Since
$d_f$ cannot be less than 1, this result was interpreted as a
strong indication that $\theta _S = 1/2$.

In this work we will demonstrate that Eqn. (1) actually {\it does}
work for both continuous and quantized distributions in 2D.  The
actual behavior of the QDE probability distributions under
finite-size scaling turns out to be more subtle than what has been
assumed until recently.\cite{Fis05}  As pointed out by Wang,
Harrington and Preskill,\cite{WHP03} domain walls of zero energy
which cross the entire sample play a special role when the energy
is quantized.

We will analyze data for the domain wall energy, $E_{dw}$, and the
domain wall entropy, $S_{dw}$, for the ground states (GS) of 2D
Ising spin glasses obtained using methods from earlier
work.\cite{SK93,SK94,LC01,Fis05}  We will demonstrate that for $L
\times L$ square lattices the Edwards-Anderson\cite{EA75} (EA)
model with a $\pm J$ bond distribution has a strong correlation
between $E_{dw}$ and $S_{dw}$ for the GS domain walls.  Because of
this correlation, we will need to treat domain walls of different
energies as distinct classes, whose entropies scale in different
ways as $L$ is increased.  We will find that the scaling parameter
identified by Saul and Kardar\cite{SK93,SK94} is the one
associated with domain walls having $E_{dw} = 0$.  It is not,
however, the one which controls the dominant behavior for large
$L$.

\section{THE MODEL}

The Hamiltonian of the EA model for Ising spins is
\begin{equation}
  H = - \sum_{\langle ij \rangle} J_{ij} {S}_{i} {S}_{j}   \, ,
\end{equation}
where each spin ${S}_{i}$ is a dynamical variable which has two
allowed states, +1 and -1.  The $\langle ij \rangle$ indicates a
sum over nearest neighbors on a simple square lattice of size $L
\times L$.  We choose each bond $J_{ij}$ to be an independent
identically distributed quenched random variable, with the
probability distribution
\begin{equation}
  P ( J_{ij} ) = x \delta (J_{ij} + 1)~+~(1 - x) \delta (J_{ij} -
  1)   \, ,
\end{equation}
so that we actually set $J = 1$, as usual.  Thus $x$ is the
concentration of antiferromagnetic bonds, and $( 1 - x )$ is the
concentration of ferromagnetic bonds.

The data analyzed here used an ensemble in which, for a given value
of $x$, every $L \times L$ random lattice sample had exactly $(1 -
x) L^2$ positive bonds and $x L^2$ negative bonds. Details of the
methods used to calculate the GS energies and the numbers of GS have
been described earlier,\cite{LC01} and the data analyzed here were
used before to obtain other properties of the model.\cite{Fis05}

\section{GROUND STATE DOMAIN WALLS}

The GS entropy is defined as the natural logarithm of the number of
ground states. For each sample the GS energy and GS entropy were
calculated for the four combinations of periodic (P) and antiperiodic
(A) toroidal boundary conditions along each of the two axes of the
square lattice.\cite{LC01}  We will refer to these as PP, PA, AP and
AA.  In the spin-glass region of the phase diagram, the variation of
the sample properties for changes of the boundary conditions is small
compared to the variation between different samples of the same
size,\cite{SK94} except when $x$ is close to the ferromagnetic phase
boundary and the ferromagnetic correlation length becomes comparable
to $L$.

We define domain walls for the spin glass as it was done in the
seminal work of McMillan.\cite{McM84b}  We look at differences
between two samples with the same set of bonds, and the same
boundary conditions in one direction, but different boundary
conditions in the other direction.  Thus, for each set of bonds we
obtain domain wall data from the four pairs (PP,PA), (PP,AP),
(AA,PA) and (AA,AP).  The reader should remember that the term
``domain wall", as used in this work, refers only to this
procedure. Saul and Kardar\cite{SK93,SK94} follow the same
procedure used in this work, but use the term ``defect" instead of
``domain wall".

It is important to realize that the meaning of a domain wall is
very different when the GS entropy is positive, as in the model we
study here, as compared to the standard case of a doubly
degenerate ground state.  In the standard case one can identify a
line of bonds which forms a boundary between regions of spins
belonging to the two different ground states.  It is not possible
to do that when there are many ground states.  Despite this, we
continue to use the term ``domain wall".

Due to the $P ( J_{ij} )$ we have chosen and the fact that we have only
used even values of $L$, the energy is always a multiple of 4.  Thus,
the energy difference, $E_{dw}$, for any pair must also be a multiple of
4.  The sign of $E_{dw}$ for a pair is essentially arbitrary for the
values of $x$ we will study here, which are deep in the spin-glass region
of the phase diagram.  Thus we can, without loss of generality, choose
all of the domain-wall energies to be non-negative.  For $x = 0.25$ and
$x = 0.5$, it turns out, crudely speaking, that about 75\% of the time we
find $E_{dw} = 0$, and 25\% of the time $E_{dw} = 4$.  For a given $L$,
the fraction of $E_{dw} = 0$ domain walls, $f_0$, is about 3\% higher at
$x = 0.5$ than at $x = 0.25$.  It is interesting to note that Wang,
Harrington and Preskill\cite{WHP03} use an analytical argument to predict
that $f_0$ is approximately 0.75, independent of $x$, in the spin-glass
regime.  The value of $f_0$ tends to increase slightly as $L$ increases,
but our statistical errors are too large for a good estimate of the
finite-size scaling behavior to be made.  Our results for $x = 0.5$ are
consistent with the results of Amoruso {\it et al.}\cite{AMMP03}

Our calculated statistics for these values of $x$, as a function of $L$,
are given in Table I and Table II.  Estimating the statistical
uncertainties in these numbers is not trivial, due to the fact that the
values of $E_{dw}$ obtained from the same set of bonds with the four
different pairs of boundary conditions are not statistically
independent.  An upper bound on the statistical uncertainties is
obtained by counting the number of samples, rather than the number of
pairs.

No domain walls with energies greater than 8 were observed at any $L$
for these values of $x$.  This, however, does not have much fundamental
significance.  The probability distribution for $E_{dw}$ is also a
function of the aspect ratio of the lattice.\cite{HBCMY02}

\begin{table}
\caption{\label{tab:table1}Domain wall energy statistics for $x =
0.25$.  $N_{AF}$ is the number of negative bonds in each sample,
and $N_{sam}$ is the number of distinct bond configurations
studied for a given $L$ and $N_{AF}$.  $n_i$ is the number of
domain walls of each type having $E_{dw} = i$, and $f_i = n_i / (4
N_{sam})$.}
\begin{ruledtabular}
\begin{tabular}{rrrrrrcc}
 $L$ & $N_{AF}$ & $N_{sam}$ & $n_0$ & $n_4$ & $n_8$ & $f_0$ & $f_4$\\
\hline
  6 & 18& 200 & 556 & 242 & 2 & 0.695 & 0.302\\
  8 & 32& 200 & 584 & 214 & 2 & 0.730 & 0.268\\
 10 & 50& 200 & 587 & 212 & 1 & 0.734 & 0.265\\
 12 & 72& 200 & 577 & 220 & 3 & 0.721 & 0.275\\
 14 & 98& 200 & 579 & 218 & 3 & 0.724 & 0.272\\
 16 &128& 200 & 573 & 226 & 1 & 0.716 & 0.282\\
 18 &162& 133 & 385 & 146 & 1 & 0.724 & 0.274\\
 20 &200& 200 & 598 & 200 & 2 & 0.748 & 0.250\\
\end{tabular}
\end{ruledtabular}
\end{table}

\begin{table}
\caption{\label{tab:table2}Domain wall energy statistics for $x =
0.5$.  Column labels as in Table I.}
\begin{ruledtabular}
\begin{tabular}{rrrrrrcc}
 $L$ & $N_{AF}$ & $N_{sam}$ & $n_0$ & $n_4$ & $n_8$ & $f_0$ & $f_4$\\
\hline
  6 & 36& 400 & 1158 & 442 & 0 & 0.724 & 0.276\\
  8 & 64& 400 & 1159 & 440 & 1 & 0.724 & 0.275\\
 10 &100& 400 & 1210 & 388 & 2 & 0.756 & 0.242\\
 12 &144& 400 & 1226 & 374 & 0 & 0.766 & 0.234\\
 14 &196& 400 & 1189 & 410 & 1 & 0.743 & 0.256\\
 16 &256& 400 & 1214 & 384 & 2 & 0.759 & 0.240\\
 18 &324& 400 & 1234 & 366 & 0 & 0.771 & 0.229\\
 20 &400& 238 &  737 & 214 & 1 & 0.774 & 0.225\\
\end{tabular}
\end{ruledtabular}
\end{table}

The domain wall entropy, $S_{dw}$, is defined, by analogy to
$E_{dw}$, to be the difference in the GS entropy when the boundary
condition is changed along one direction from P to A, with the
boundary condition in the other direction remaining fixed.  The
probability distribution of $S_{dw}$ for the cases where
$E_{dw} = 0$ should be symmetric about 0, and our statistics are
consistent with this.  As shown by Saul and Kardar,\cite{SK93,SK94}
the variance of $S_{dw}$ when $E_{dw} = 0$ increases with $L$ in
approximately a linear fashion.  This is illustrated in Fig. 1,
for both $x=0.25$ and $x=0.5$.  The calculated skewness of these
essentially symmetric distributions is, naturally, consistent with
zero, but their kurtosis is not.  For $x=0.25$ the calculated
kurtosis for different values of $L$ varies between 1.0 and 2.5,
with no apparent trend in the $L$ dependence.  For $x=0.5$, where
the number of samples is about twice as large, the kurtosis varies
between 1.35 and 2.05, again without any apparent trend with $L$.
If we average over $L$, we obtain $1.58 \pm 0.18$ for $x = 0.25$
and $1.70 \pm 0.10$ for $x = 0.5$.  From this we conclude that the
$E_{dw} = 0$ distributions are not Gaussian, and that they
probably do not become Gaussian even as $L$ becomes large.  The
results are consistent with a probability distribution which has
a kurtosis of about 1.7, independent of $L$ and $x$.  Since the
model is critical at $T = 0$ for both values of $x$, this result
is not surprising.

\begin{figure}
\includegraphics[width=3.4in]{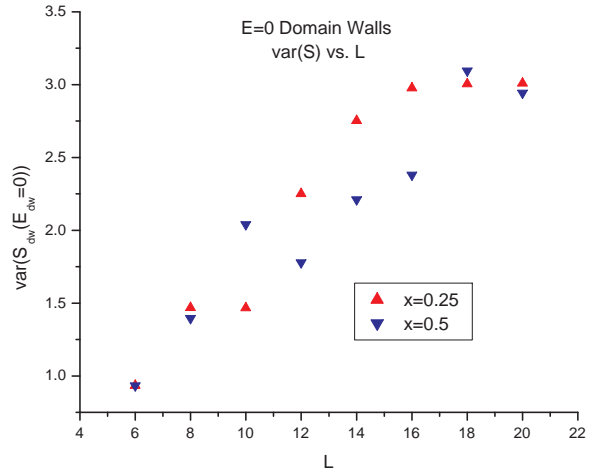}
\caption{\label{Fig.1} Variance of the entropy vs. $L$ for the
$E_{dw} = 0$ domain walls, for the cases $x = 0.25$ and $x =
0.5$.}
\end{figure}

\begin{figure}
\includegraphics[width=3.4in]{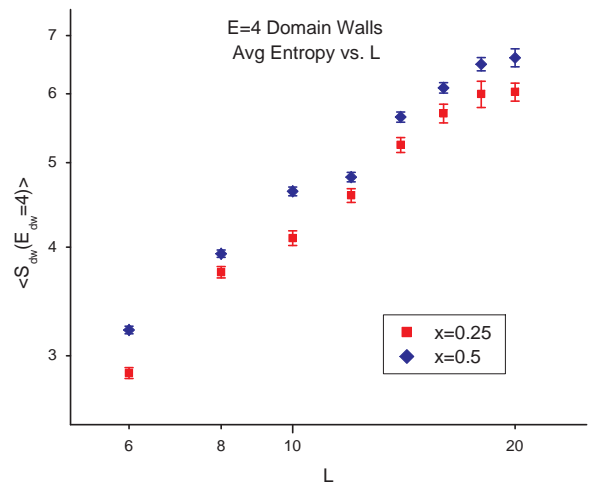}
\caption{\label{Fig.2} Average entropy vs. $L$
for the $E_{dw} = 4$ domain walls, for the cases $x = 0.25$
and $x = 0.5$, log-log plot.  The error bars indicate 1
$\sigma$, and the statistical errors are larger for $x = 0.25$
because the numbers of samples are smaller.}
\end{figure}

When $E_{dw}$ is not zero, the relative signs of $E_{dw}$ and $S_{dw}$
are not arbitrary.  Having chosen $E_{dw}$ to be nonnegative, we
then find that, when $E_{dw}$ is positive, it turns out that $S_{dw}$
is almost always positive.  In Fig. 2 we show the behavior of the
average value of $S_{dw}$ for the cases where $E_{dw} = 4$, as a
function of $L$.  We see that for $E_{dw} = 4$, the average value
of $S_{dw} ( L )$ grows approximately as $L^{0.63}$, for both
values of $x$.  More precisely, a least-squares fit to the form
\begin{equation}
S_{dw} ( L ) = A L^{\theta_S}  \,
\end{equation}
gives
\begin{equation}
\theta_S = 0.639 \pm 0.036  \,
\end{equation}
for $x = 0.25$, and
\begin{equation}
\theta_S = 0.628 \pm 0.027  \,
\end{equation}
for $x = 0.5$.  These numbers should be compared to the prediction
of droplet theory,\cite{FH88} Eqn. (2).  Because of the large GS
degeneracy in the $\pm J$ Ising spin glass, one does not know how
to compute $d_f$ directly for this model.

For the Ising spin glass with a Gaussian bond distribution,
however, this calculation in 2D is straightforward, and the
result\cite{BM87,Mid01,HY02} is $d_f = 1.27 \pm 0.01$.  It would
be desirable to improve the accuracy of our estimate of $\theta_S$
sufficiently to demonstrate that $\theta_S$ is really different
for the CDE and the QDE, as predicted by Eqn. (1).  The predicted
numerical difference is small, however, so this will be difficult.

The skewness and kurtosis of the $S_{dw}$ distributions for
$E_{dw} = 4$ are both small, in contrast to the $E_{dw} = 0$ case.
It is likely that these distributions become Gaussian in the large
$L$ limit.

\begin{figure}
\includegraphics[width=3.4in]{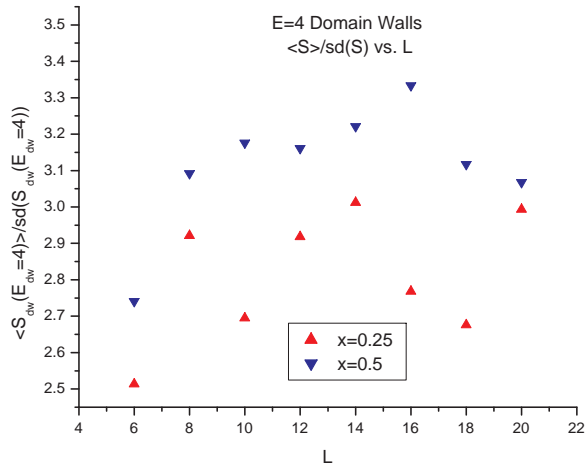}
\caption{\label{Fig.3} Average entropy divided by its
standard deviation vs. $L$ for the $E_{dw} = 4$ domain
walls, for the cases $x = 0.25$ and $x = 0.5$.}
\end{figure}

The result of Saul and Kardar,\cite{SK93,SK94} obtained by looking
at the distribution of $S_{dw}$ for all values of $E_{dw}$
combined, was $\theta_S = 0.49 \pm 0.02$.  To obtain this
exponent, Saul and Kardar fit the data at small values of
$S_{dw}$.  For practical purposes, this part of the data belongs
to the $E_{dw} = 0$ component.  It is clear that if our
extrapolations from small $L$ are correct, however, eventually, at
large enough $L$, the width of the total $S_{dw}$ distribution
will be dominated by the $E_{dw} = 4$ component. Although they did
not identify it, the peak in the probability distribution of
$S_{dw}$ due to the $E_{dw} = 4$ component is clearly visible in
Fig. 19 of the 1994 paper of Saul and Kardar.\cite{SK94}

In Fig. 3 we show the values of the ratio of average value of the
$S ( L )$ distribution for $E_{dw} = 4$ divided by the rms width of
the distribution.  Except for $L = 6$, where the ratios are smaller,
we see that this ratio is about 2.9 for $x = 0.25$, and about 3.2
for $x = 0.5$.  Thus this characteristic of the distributions seems
to depend on $x$.  However, it seems likely that the ratio is
actually $L$-dependent, and that the large $L$ limit is universal,
independent of $x$.

\section{DISCUSSION}

What we have learned is that in this model there are two distinct
classes of domain walls, the $E_{dw} = 0$ domain walls and the
$E_{dw} > 0$ domain walls.  For $x = 0.25$ or $x = 0.5$, the only
type of $E_{dw} > 0$ domain walls which we need to worry about are
the $E_{dw} = 4$ ones.  If we wanted to study values of $x$ of
0.15 or less,\cite{AH04} or values of the aspect ratio less than
one,\cite{FH05} then we would need to consider larger values of
$E_{dw}$.  It should be expected that these are qualitatively
similar to the $E_{dw} = 4$ case, as explained below.

As we have seen, the $E_{dw} = 4$ domain walls behave in a way
which is essentially consistent with the predictions of the droplet
model, but the $E_{dw} = 0$ domain walls do not.  This difference
in behavior cannot be understood within the droplet model.  It
must be related to the special feature of the model caused by the
quantization of the energy spectrum, which is the positive GS
entropy.  Another fact which we need to use is the recently
discovered result\cite{Fis05} that the average GS entropy is an
increasing function of the GS energy.

For an $E_{dw} > 0$ domain wall, a contribution to $S_{dw}$ comes
from this shift in the average GS entropy with the shift in the
GS energy.  What remains to be understood is why $S_{dw}$ should
scale with increasing $L$ in the way predicted by the droplet
model.  The conventional derivation of the droplet model\cite{FH86}
appears to use the assumption that the GS is unique, up to a
reversal of the entire state, in an essential way.  The
extension of the droplet model to the more general case was
given by Fisher and Huse.\cite{FH88}

As long as $E_{dw} > 0$, the two boundary conditions which we are
comparing are not on an equal footing.  As Wang, Harrington and
Preskill\cite{WHP03} express the situation, the $E_{dw} > 0$ domain
wall does not destroy the topological long-range order.  However, in
the $E_{dw} = 0$ case the two boundary conditions are on an equal
footing, and the topological order is destroyed.  Therefore the
$E_{dw} = 0$ class of domain walls can be expected to behave in a
special way, which differs from the prediction of the droplet model.

It is well established that this model displays large corrections
to finite-size scaling.\cite{CBM03}  The reader may object that
nothing reliable about the large $L$ behavior can be learned from
data which is restricted to $L \le 20$.  However, our results
indicate that the reason for the large corrections to scaling is
that the differences in the scaling exponents for different
classes of domain walls is small. When each class is analyzed
separately, this effect is eliminated.  Thus, going out to only $L
= 20$ is not as bad as it might appear at first.  A more
significant limitation in our work is that the number of samples
we have data for is much smaller than we would like.

We also wish to point out that there is a natural similarity
between the $E_{dw} = 0$ domain walls in the 2D Ising spin glass
with $\pm J$ bonds and the large-scale low energy excitations
which have been found\cite{KM00,PY00} in the three-dimensional
version of the same model.  A similar, and possibly related
large-scale low energy excitation has also been seen in the 3D
random-anisotropy XY model.\cite{Fis95}  Similar ideas have also
been discussed by Hatano and Gubernatis.\cite{HG02}

The domain-wall renormalization group of McMillan\cite{McM84a} is
based on the idea that we are studying an effective coupling
constant which is changing with $L$.  For the case of a continuous
distribution of bonds, we can use the energy as the coupling
constant. For the quantized energy case, what we should really do
is a slight generalization of this idea. We should think of the
coupling constant as the free energy at some infinitesimal
temperature.  When we do this, the entropy contributes to the
coupling constant.  And as we have seen, the distribution of
$E_{dw}$ rapidly becomes essentially independent of $L$ as $L$
becomes large.  Under these conditions, it becomes natural to
treat each value of $E_{dw}$ as a separate class, representing a
different coupling constant.

Therefore, since $S_{dw}$ increases as a positive power of $L$ for
$E_{dw} = 4$, and presumably for all other (finite) positive
values, this coupling constant must eventually be controlled by
$S_{dw}$.  And this coupling constant is also the one described by
the droplet model.

The $E_{dw} = 0$ class behaves differently, and therefore
represents a different coupling constant.  It is natural to wonder
if topological long-range order can be related to replica-symmetry
breaking, and if the $E_{dw} = 0$ domain walls can be described by
the replica-symmetry breaking theory.  We will not attempt to do
this here.

The argument that all classes other than the $E_{dw} = 0$ class
behave in the same way should certainly be checked.  The way to do
this would be to do calculations with odd values of
$L$.\cite{BM86} In this way we would obtain results for the
$E_{dw} = 2$ class, and possibly also for the $E_{dw} = 6$ class.

The computing effort used to obtain the results described here was
rather modest by current standards.  The algorithm\cite{LC01} is
capable of generating data for greater numbers of samples out to
somewhat larger values of $L$ (perhaps to $L = 32$) with a
reasonable effort.  The method of Lukic {\it et al.}\cite{LGMMR04}
appears to be more powerful, and has been used by them to generate
energy-entropy statistics for ground states of this model out to
$L = 50$.

\section{SUMMARY}

We have studied the statistics of domain walls for ground states
of the 2D Ising spin glass with +1 and -1 bonds for $L \times L$
square lattices with $L \le 20$, and $x$ = 0.25 and 0.5, where $x$
is the fraction of negative bonds, using periodic and/or
antiperiodic boundary conditions. Under these conditions, almost
all domain walls have an energy $E_{dw}$ equal to 0 or 4.  The
probability distribution of the entropy $S_{dw}$ is found to
depend strongly on $E_{dw}$.  The results for $S_{dw}$ when
$E_{dw} = 4$ agree with the prediction of Amoruso, Hartmann,
Hastings and Moore,\cite{AHHM06} Eqn. (1). Our results for
$S_{dw}$ when $E_{dw} = 0$ agree with those of Saul and
Kardar,\cite{SK93,SK94} but in addition we find that the
distributions are not Gaussian in that case, even in the limit of
large $L$. Due to the special role of the $E_{dw} = 0$ domain
walls, we can understand the difference between the scaling
exponent found by Saul and Kardar and the prediction of Eqn. (1).

\begin{acknowledgments}
The author thanks S. N. Coppersmith for generously providing all of
the raw data analyzed in this work.  He is grateful to S. L. Sondhi,
A. K. Hartmann, D. F. M. Haldane and D. A. Huse for helpful discussions,
and to Princeton University for providing use of facilities.

\end{acknowledgments}



\end{document}